%% file: gw_paper.tex
\documentclass[twocolumn]{aastex631}

\makeatletter
\let\frontmatter@title@above=\relax

\newcommand{\Msun}{\ensuremath{M_\odot}\xspace}

\definecolor{chmagenta}{rgb}{0.54, 0.17, 0.88}

\newcommand{\CIERA}{\affiliation{Center for Interdisciplinary Exploration and Research in Astrophysics (CIERA), Northwestern University, 1800 Sherman Ave, Evanston, IL 60201, USA}}
\newcommand{\Princeton}{\affiliation{Department of Physics, Princeton University, Princeton, NJ 08544, USA}}

\usepackage{acronym}
\usepackage{amsmath}
\usepackage{xspace}

\graphicspath{{./}{figures/}}
\input{acronyms}
\input{macros}

\begin{document}

\title{Characterizing Compact-object Binaries in the Lower Mass Gap with Gravitational Waves}

\author[0009-0003-2137-8213]{Jessica Cotturone}\email{jac18@illinois.edu}
\affiliation{Augustana College, 639 38th St, Rock Island, IL 61201, USA}
\CIERA
\affiliation{Siebel School of Computing and Data Science, University of Illinois Urbana-Champaign, 201 N. Goodwin Ave, Urbana, IL 61801, USA}

\author[0000-0002-0147-0835]{Michael Zevin}\email{mzevin@adlerplanetarium.org}
\CIERA
\affiliation{Department of Astronomy, Adler Planetarium, 1300 S. Lake Shore Drive, Chicago, IL 60605, USA}
\affiliation{NSF-Simons AI Institute for the Sky (SkAI), 172 E. Chestnut St., Chicago, IL 60611, USA}

\author[0000-0001-7616-7366]{Sylvia Biscoveanu}\email{sbisco@princeton.edu}\thanks{NASA Einstein Fellow}
\CIERA\Princeton

\begin{abstract}

The source binary of the gravitational-wave (GW) event GW230529, detected at the beginning of the fourth LIGO-Virgo-KAGRA observing run, was inferred to consist of a NS and a compact object in the lower mass gap, a purported gap between the most massive NSs ($\sim 3$~\Msun) and least massive black holes (BHs; $\sim 5$~\Msun) based on compact-object observations in the Milky Way. While the nature of the mass-gap object could not be determined from the GW data alone for this event, definitively distinguishing whether this object is a NS or BH would have profound implications for the NS equation of state, supernova physics, and multimessenger astronomy.
In this work, we perform parameter estimation on a suite of simulated GW systems with parameters similar to those of the GW230529 source binary to investigate whether the ambiguity in the physical nature of the source is a generic result for such systems.
We vary the intrinsic properties of the simulated systems, the detector noise properties, the signal-to-noise ratios (SNRs), and the waveform model used in recovery. 
We find that the low SNR of GW230529 is the key reason for the ambiguity in determining whether the mass of the primary object in the binary is consistent with a low-mass BH or a high-mass NS, and thus the priors used for the masses and spins have a significant impact on the posterior distribution, which is a generic result for low-SNR events.
The inclusion of tidal effects in the waveform model also contributes to the observed degeneracies in the posteriors, since the statistical uncertainties in analyses of GW events like GW230529 are larger for waveform models including tidal effects.
We show that the future observation of such a system with a higher SNR ($\sim 30$) would increase the precision of the mass measurements sufficiently to allow us to determine the nature of the mass-gap object.
\end{abstract}


\section{Introduction} \label{sec:intro}

During the first three observing runs of the international \ac{GW} detector network, the \ac{LVK} Collaboration reported nearly 100 significant compact binary coalescence candidates~\citep{KAGRA:2021vkt}.
These observations include all of the variations of \ac{BH} and \ac{NS} mergers: \ac{BBH}, \ac{BNS}, and \ac{NSBH} systems.
The ongoing \ac{LVK} \ac{O4} began in 2023 May after the completion of advanced technical upgrades to improve the sensitivity of the detectors~\citep{KAGRA:2013rdx, TheLIGOScientific:2014jea, TheVirgo:2014hva, Aso:2013eba, Somiya:2011np, KAGRA:2020tym, Cahillane:2022pqm}.
The first significant candidate observed in this run was GW$230529\_181500$ (hereafter referred to as GW230529), which was detected by LIGO Livingston with a \ac{SNR} of $\approx\originalEventSNR$. The other interferometers in the \ac{LVK} network were either offline at the time of the event or not sensitive enough to detect the signal~\citep{LIGOScientific:2024elc}.

The primary object in GW230529 was inferred to have a mass within the ``lower mass gap,'' defined as the range in the compact-object mass distribution between the heaviest \acp{NS} ($\sim3$~\Msun; \citealt{Rhoades:1974fn,Kalogera:1996ci}) and the lightest \acp{BH} ($\sim5$~\Msun; \citealt{Bailyn:1997xt,Ozel:2010su,Farr:2010tu}) observed in the Milky Way.
Theoretical models explain this observational gap between \ac{NS} and \ac{BH} masses in terms of the timescale for instability growth in the core collapse of the massive stellar progenitors of these compact objects. A rapid timescale for instability growth suppresses fallback accretion, leading to a gap in the compact-object mass distribution~\citep{Fryer:1999ht,Fryer:2011cx,Belczynski:2011bn}. 
The lower end of this gap depends on the maximum mass of an astrophysical \ac{NS}, which cannot exceed the maximum mass imposed by the unknown \ac{NS} \ac{EOS}. 
Although recent analyses that consider electromagnetic and \ac{GW} observations constrain the maximum mass of a nonspinning \ac{NS} to $M_{\mathrm{TOV}}\sim2.0\text{--}2.7$~\Msun and find no evidence for a different astrophysical maximum mass~\citep{Golomb:2024lds}, the mass of a spinning \ac{NS} can exceed $M_{\mathrm{TOV}}$.
Meanwhile, the upper bound of the lower-mass-gap range, $\sim 5$~\Msun, arises from a paucity of observations of Milky Way \acp{BH} below this mass.

Several potential mass-gap objects have been identified in the past, primarily through electromagnetic observations. For instance, \cite{Thompson:2018ycv} and \cite{Jayasinghe:2021uqb} report on noninteracting binary systems each containing an unseen compact-object companion to a red giant, where the masses of the companion in the two systems are inferred to be $3.3^{+2.8}_{-0.7}$~\Msun and $3.04\pm0.06$~\Msun, respectively. Another compact object mass-gap candidate, with a mass between 2.09 and 2.71~\Msun, was recently identified through radio pulsar observations~\citep{Barr:2024wwl}. 
\ac{GW} observations such as GW190814 have also provided potential evidence for mass-gap objects; however, this is not significant enough to rule out the existence of the lower mass gap between \acp{NS} and \acp{BH}~\citep{LIGOScientific:2020zkf, KAGRA:2021duu}.

The discovery of GW230529~\citep{LIGOScientific:2024elc} provides the strongest evidence to date for an object in the lower mass gap since the measurement of the primary mass is constrained between 2.4 and 4.4~\Msun (all measurements are reported as symmetric 90\% credible intervals around the median of the marginalized posterior distribution).
Analyzing this system can therefore reveal more about the properties, formation channels, and merger rates for objects in the lower mass gap. 
Although several alternative scenarios have been proposed for the formation of the GW230529 source~\citep{Afroz:2024fzp, Afroz:2025efn, Zhu:2023nhy, Xing:2024ydg, Qin:2024ojw, Janquart:2024ztv, Ye:2024wqj, Zhu:2024cvt, Huang:2024wse, Mahapatra:2025agb}, we here assume that the components of the system formed from standard stellar core collapse.
However, given the uncertainty in the mass measurement and lack of measurable tidal information in the signal, it remains unclear whether the primary object in GW230529 is a low-mass \ac{BH} or a massive \ac{NS}.
The ability to classify the compact object is limited by gaps in our understanding of the lower mass gap and the \ac{NS} \ac{EOS}, as well as ambiguities in the measured mass posterior distributions.
Given current constraints on the \ac{EOS}, the most likely interpretation is a mass-gap \ac{BH} merging with a \ac{NS}; however, we cannot exclude the possibility that the two compact objects are heavy ($\gtrsim 2$~\Msun), near-equal-mass \acp{NS}~\citep{LIGOScientific:2024elc}.
Both hypotheses would have major astrophysical implications.

If the primary object in GW230529 is a stellar-mass \ac{BH}, it would be one of the smallest \acp{BH} observed (see also \citealt{LIGOScientific:2020zkf}). 
This would provide definitive evidence that a population of compact objects exists within the lower mass gap.
It would also be the most equal-mass \ac{NSBH} yet observed, which enhances the probability that tidal disruption could occur in such systems and produce remnant material capable of powering an electromagnetic counterpart such as a kilonova \citep{Lattimer:1974slx, Li:1998bw, Tanaka:2013ana, Tanaka:2013ixa, Fernandez:2016sbf, Kawaguchi:2016ana} or a gamma-ray burst \citep{Mochkovitch:2021prz, Janka:1999qu, Paschalidis:2014qra, Shapiro:2017cny, Ruiz:2018wah}.
There were no significant counterpart candidates observed for GW230529~\citep{icecube2023, karambelkar2023, lipunov2023, longo2023, lesage2023, savchenko2023, sugita2023a, sugita2023b, waratkar2023}, but effective electromagnetic follow-up efforts were limited because the event was only observed with one detector and therefore not well localized on the sky.
If instead GW230529 were a merger between two heavy \acp{NS}, it would be the heaviest \ac{BNS} system observed so far. 
This observation would have a profound impact on the \ac{NS} \ac{EOS}, as current inferences strongly disfavor \acp{NS} with masses as large as the primary object of GW230529 would require in a \ac{BNS} scenario~\citep[e.g.,][]{Legred:2021hdx, Miller:2021qha, Raaijmakers:2021uju}.

Many previous works have investigated the precision with which the parameters of \ac{NSBH} systems can be measured, as well as the roles of various sources of statistical and systematic uncertainty~\citep{vanderSluys:2007st, Cho:2012ed, Hannam:2013uu, OShaughnessy:2014shr, Chatziioannou:2014coa, Vitale:2014mka, Kumar:2016zlj, Huang:2020pba}.
These works generally find that the absolute precision of the mass measurements improves with decreasing mass, and that spin precession can help break the mass-ratio--spin degeneracy that limits the ability to accurately measure the component masses of low-mass binaries~\citep{Apostolatos:1994mx}.
Other studies have focused on distinguishing \ac{NSBH} systems from \acp{BNS} using the tidal deformability parameter and the \ac{NS} \ac{EOS} both for individual events~\citep{Chen:2020fzm, Datta:2020gem} and the population as a whole~\citep{Fasano:2020eum, Essick:2020ghc}, despite the weak signatures that tidal effects imprint on the waveforms of massive \ac{BNS} mergers, which make them difficult to distinguish from \acp{BBH} of the same total mass~\citep{Yang:2017gfb, Tsokaros:2019lnx}.
Furthermore, it is easier to determine the nature of compact-object binaries including sub-solar-mass components than those including lower-mass-gap objects, using both mass~\citep{Wolfe:2023yuu} and tidal deformability measurements~\citep{Golomb:2024mmt}, as the predicted tidal deformability generally increases (and hence becomes more measurable) with decreasing \ac{NS} mass.
\cite{Littenberg:2015tpa} focused specifically on the ability to distinguish the nature of compact-object mergers including a lower-mass-gap component, finding that only \acp{NS} with masses $< 1.5~\Msun$ and \acp{BH} with masses $> 6~\Msun$ can be confidently identified.

In this work, we seek to determine how well we can characterize GW230529-like systems by performing parameter estimation on simulated signals.
Like \cite{Littenberg:2015tpa}, we focus on the measurement of the component masses rather than the tidal deformability parameters when classifying the compact objects in our simulations, but are additionally interested in the source of the multimodality in the posteriors of GW230529 that leads to ambiguity in its source classification.
By exploring mass-gap signals through this simulation study, we can better assess our ability to characterize compact binary systems, including future \ac{GW} events with properties similar to GW230529.
We find that it is difficult to distinguish between the massive \ac{BNS} versus \ac{NS} and low-mass \ac{BH} hypotheses for systems similar to GW230529 due to the large statistical uncertainties inferred for these events at low \ac{SNR}, especially when potential tidal effects are taken into account.
However, we show that higher \acp{SNR} in future \ac{GW} detections of mass-gap systems will make it easier to characterize these systems and to differentiate between these two hypotheses.

The format of this paper is as follows. 
Section~\ref{sec:methods} describes the parameter estimation methods we use to analyze simulated \ac{GW} signals and the intrinsic binary parameters chosen as the baseline for much of our study. 
Section~\ref{sec:intrinsic_params} explains the setup and results from simulating \acp{GW} with different intrinsic parameters. This section also includes a more general analysis covering intrinsic mass and spin parameters beyond those corresponding to GW230529-like systems to determine how well the properties of different lower-mass-gap binary systems can be measured. 
In Section~\ref{sec:noise_realizations}, we describe the impact of adding detector noise to the simulated signals, and in Section~\ref{sec:increasing_snr}, we analyze how increasing the \ac{SNR} affects the precision of our measurements. 
We next explore parameter estimation with different waveform models, and compare the results in Section~\ref{sec:waveform_models}. Finally, Section~\ref{sec:conclusions} summarizes the main questions addressed in this study and the answers found by our simulation campaign.

\section{Parameter Estimation Methods} \label{sec:methods}

Bayesian inference techniques can be used to infer the properties of a compact binary system from \ac{GW} strain data.
For our study, we perform parameter estimation on simulated \ac{GW} signals with properties characteristic of \ac{NSBH} systems with lower-mass-gap \acp{BH} to investigate how the true parameters of the system, specific noise realization, \ac{SNR}, and waveform model affect uncertainties in the results. 
We utilize the nested sampler \textsc{Dynesty}~\citep{Skilling:2006gxv, Speagle:2019ivv} to explore the parameter space and sample posterior probability distributions for the binary parameters, characterizing the simulated signals using the \textsc{Bilby} inference library~\citep{Ashton:2018jfp, Romero-Shaw:2020owr}.
The posterior distribution for the parameters $\boldsymbol{\theta}$ given the observation of data $d$ is given by Bayes' theorem:
\begin{equation}
p(\boldsymbol{\theta}|d) = \frac{\mathcal{L}(d|\boldsymbol{\theta})\pi(\boldsymbol{\theta})}{\mathcal{Z}}
\end{equation}
where $\mathcal{L}(d|\boldsymbol{\theta})$ is the likelihood of observing the data $d$ for a set of parameter values $\boldsymbol{\theta}$, $\pi(\boldsymbol{\theta})$ is the prior probability distribution for the binary parameters, and the normalization factor $\mathcal{Z}$ is the evidence, or marginalized likelihood:
\begin{equation}
\mathcal{Z} = \int{d\theta\mathcal{L}(d|\boldsymbol{\theta})\pi(\boldsymbol{\theta})}.
\end{equation}
For \ac{GW} data analysis, we typically assume the data are stationary, Gaussian, and characterized by a known power spectral density (PSD), $S$, so the likelihood in a single interferometer is~\citep[e.g.,][]{Romano:2016dpx}
\begin{equation}
\mathcal{L}(d|\boldsymbol{\theta}) \propto \exp{\bigg(-\sum_k{\frac{2|d_k - h_k(\boldsymbol{\theta})|^2}{TS_k}}\bigg)},
\end{equation}
where $h(\boldsymbol{\theta})$ is the gravitational waveform that depends on the binary parameters, $T$ is the duration of the analyzed data, and the subscript $k$ indicates the frequency. 
We choose the duration, frequency range, and PSD to be the same as in the GW230529 \ac{LVK} analysis~\citep{LIGOScientific:2024elc, ligo_scientific_collaboration_2024_10845779}.

Consistent with \ac{LVK} convention, the priors we use for our analyses are uniform in detector-frame component masses, isotropic in spin direction, and uniform in spin magnitude, with the restrictions that $\chi_1\le0.99$ and $\chi_2\le0.05$. 
This low secondary-spin prior is astrophysically motivated by observations of Galactic \acp{BNS} that merge within a Hubble time~\citep{Burgay:2003jj, Stovall:2018ouw}.
We sample in detector-frame chirp mass and mass ratio, with prior ranges equivalent to those used in the LVK analysis of GW230529 (the prior range for detector-frame chirp mass is $[\minChirpPrior~\Msun$, $\maxChirpPrior~\Msun$], and the prior range for mass ratio is $[0.125, 1]$).
Unless explicitly noted, we use a \ac{BNS} waveform model \texttt{IMRPhenomPv2\_NRTidalv2}~\citep{Dietrich:2019kaq}, which includes the effects of spin precession and allows for tidal effects in both components of the binary, for data generation and parameter estimation in all of our simulations. We also use \texttt{IMRPhenomXPHM}~\citep{Pratten:2020fqn, Garcia-Quiros:2020qpx, Pratten:2020ceb}, which models precessing \acp{BBH} and includes \ac{GW} emission from higher-order modes.
For parameter estimation with the \texttt{IMRPhenomPv2\_NRTidalv2} waveform, we use the reduced-order quadrature (or ROQ) likelihood acceleration method~\citep{Canizares:2013ywa, Smith:2016qas, Morisaki:2023kuq}.
With \texttt{IMRPhenomXPHM}, we use the multibanding likelihood method~\citep{Garcia-Quiros:2020qlt, Morisaki:2021ngj}.

The three parameters we focus on for the purposes of this study are the primary mass~($m_1$, the mass of the more massive object in the binary system), the mass ratio~($q$), and the effective inspiral spin~($\chi_{\mathrm{eff}}$). 
The mass ratio is defined as $q=m_2/m_1$ so that $0 < q \leq 1$. 
The spin parameter $\chi_{\mathrm{eff}}$ is a mass-weighted combination of the spin components along the binary's orbital angular momentum, and is defined as
\begin{equation}
\chi_{\mathrm{eff}} = \frac{{\chi}_1 \cos{\theta_1} +  {\chi}_2 q \cos{\theta_2}}{1+q},
\end{equation}
where $\chi_i$ is the magnitude of the component spin for $i=1,2,$ and $\theta_i$ is the tilt of the component spin relative to the orbital angular momentum~\citep{Damour:2001tu, Ajith:2009bn, Ajith:2011ec, Santamaria:2010yb, Purrer:2013ojf}.
While \ac{GW} observations of low-mass systems best constrain the binary chirp mass, $\mathcal{M} = (m_1 m_2)^{3/5}/(m_1 + m_2)^{1/5}$, we use the measurement of the primary mass $m_1$ to discern whether or not the primary object in GW230529 is consistent with a \ac{BH} or a \ac{NS}. 

\begin{figure}
\includegraphics[scale=0.3,angle=0]{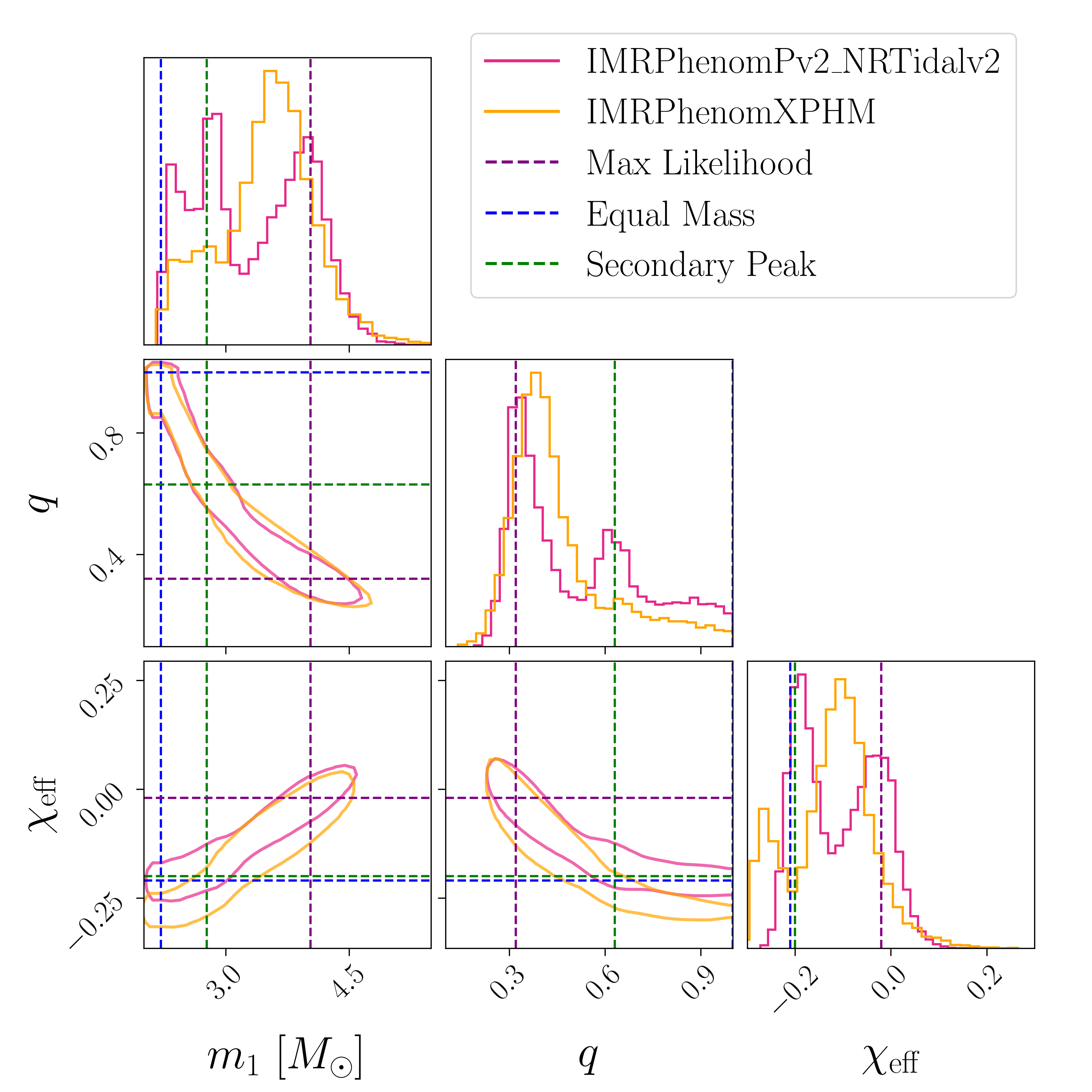} 
\caption{Posterior probability distributions of the primary mass, mass ratio, and effective inspiral spin obtained from parameter estimation of the GW230529 signal by the \ac{LVK}~\citep{ligo_scientific_collaboration_2024_10845779}. Strain data used for analysis comes from the LIGO Livingston detector. The posteriors obtained using the \texttt{IMRPhenomPv2\_NRTidalv2} (\texttt{IMRPhenomXPHM}) waveform model are shown in pink (yellow). Lines corresponding to the \textsc{Max Likelihood}, \textsc{Equal Mass}, and \textsc{Secondary Peak} samples that we use for our analysis are included for reference.}
\label{fig:gw230529_pe}
\end{figure}

The measurements of $q$ and $\chi_{\mathrm{eff}}$ are strongly correlated with the primary mass: At the same chirp mass, a system with a larger $m_1$ will have a lower $q$ (i.e., more asymmetric) and a $\chi_{\mathrm{eff}}$ value near zero, whereas a system with a smaller $m_1$ has a higher $q$ (i.e., more symmetric) and negative $\chi_{\mathrm{eff}}$.
The correlation between these three parameters and the notably broad and multimodal distributions obtained from parameter estimation on the real GW230529 data, using both the \texttt{IMRPhenomPv2\_NRTidalv2} and \texttt{IMRPhenomXPHM} waveform models, are shown in Fig.~\ref{fig:gw230529_pe}. The primary mass and mass ratio are correlated along lines of constant chirp mass, $m_{1} = \mathcal{M}(1+q)^{1/5}/q^{3/5}$, and the mass ratio and effective spin are correlated because of their degenerate effects on the 1.5PN (post Newtonian) coefficient where spin first enters the expansion of the \ac{GW} phase for a compact-object binary~\citep{Cutler:1994ys, Poisson:1995ef, Kidder:1992fr, Baird:2012cu, Ng:2018neg}: 
\begin{align}
    \psi_{1.5} &= (\pi\mathcal{M}f)^{-2/3}\psi,\\
    \psi &= \eta^{-3/5}\left[ \frac{(113-76\eta)}{128}\chi_{\mathrm{eff}} + \frac{76\delta\eta}{128}\chi_{a} - \frac{3\pi}{8}\right],
\end{align}
where $\chi_{a} = (\chi_{z,1} - \chi_{z,2})/2,\ \delta = (m_{1}-m_{2})/(m_{1}+m_{2})$, and $\eta = q/(1+q)^{2}$ is the symmetric mass ratio. For low-mass, inspiral-dominated sources like GW230529, the mass ratio $q$ and effective aligned spin are correlated along lines of constant $\psi$. This parameter is well measured and uncorrelated with mass ratio, particularly for unequal-mass systems with posterior skewness toward negative $\chi_{\mathrm{eff}}$ values like GW230529~\citep{Ng:2018neg}.

\section{Varying the Intrinsic Parameters} \label{sec:intrinsic_params}

For our initial simulations, we generate signals with true parameter values that correspond to specific samples of the posterior obtained in the \ac{LVK} \texttt{IMRPhenomPv2\_NRTidalv2} analysis of GW230529 using the low secondary-spin prior described in Section~\ref{sec:methods} ~\citep{ligo_scientific_collaboration_2024_10845779}. 
We refer to these samples as \textsc{Max Likelihood}, \textsc{Equal Mass}, and \textsc{Secondary Peak}. 
The parameter values for each of these samples are shown in Fig.~\ref{fig:gw230529_pe}.
Each of the three parameter sets we consider would have profound astrophysical implications if they represent the true properties of the GW230529 progenitor.

\textsc{Max Likelihood} corresponds to the posterior sample with the largest likelihood value, with unequal masses characteristic of a mass-gap \ac{BH} merging with a standard-mass \ac{NS}. 
If representative of the true GW230529 system, this would indicate that \acp{BH} can exist in the lower mass gap, with implications for both the supernova mechanism and the multimessenger prospects of the \ac{NSBH} population~\citep{LIGOScientific:2024elc}.

\textsc{Equal Mass} has parameters corresponding to the posterior sample with the most equal mass ratio, with component masses characteristic of two heavy \acp{NS}. If we could determine that GW230529 corresponded to such a system, this would provide additional evidence for the existence of massive \acp{NS}~\citep[e.g.,][]{Romani:2022jhd}, improving constraints on the \ac{NS} \ac{EOS}. Such a massive \ac{BNS} system would be a significant outlier relative to the Galactic population of double \acp{NS}~\citep{Farr:2010tu}, suggesting either an observational selection effect or distinct formation processes leading to differences in the mass distributions for the population observed with \acp{GW} versus the population observed electromagnetically~\citep[e.g.,][]{Galaudage:2020zst, Romero-Shaw:2020aaj, Safarzadeh:2020efa}.

\textsc{Secondary Peak} corresponds to the maximum-likelihood sample out of the subset of samples for which $m_{1}< 3~\Msun$, selected to capture the secondary mode in the original $m_{1}$ posterior. 
With a primary mass of $\approx 2.8~\Msun$, if this choice of parameters is representative of the true GW230529 system, the primary compact object would have a mass similar to the secondary of GW190814~\citep{LIGOScientific:2020zkf}, and the binary system may be an analog of the pulsar with a $\sim 2.1$--$2.7$~\Msun companion observed by MeerKAT~\citep{Barr:2024wwl}.
The key binary parameters for each chosen sample, along with the optimal \ac{SNR} in the LIGO Livingston detector, are given in Table~\ref{tab:params}.

\input{parameters_table}
\vspace{-8mm}

\subsection{GW230529-like systems}
\label{sec:sim_posterior_samples}
We initially investigate whether the primary mass can be measured well enough to discern whether the object is a \ac{BH} or \ac{NS}, based on current \ac{EOS}-based constraints on the \ac{NS} maximum mass.
To do this, we simulate systems corresponding to each set of intrinsic parameters in Table~\ref{tab:params} and perform parameter estimation on the synthetic \ac{GW} signals these systems produce.
In each simulation, we do not add detector noise to the \ac{GW} signal; we will explore the effects of simulated detector noise in Section~\ref{sec:noise_realizations}.
Since we aim to replicate the circumstances of the real observation of GW230529, we use a single-detector (LIGO Livingston--only) configuration.

\begin{figure}
\includegraphics[scale=0.3,angle=0]{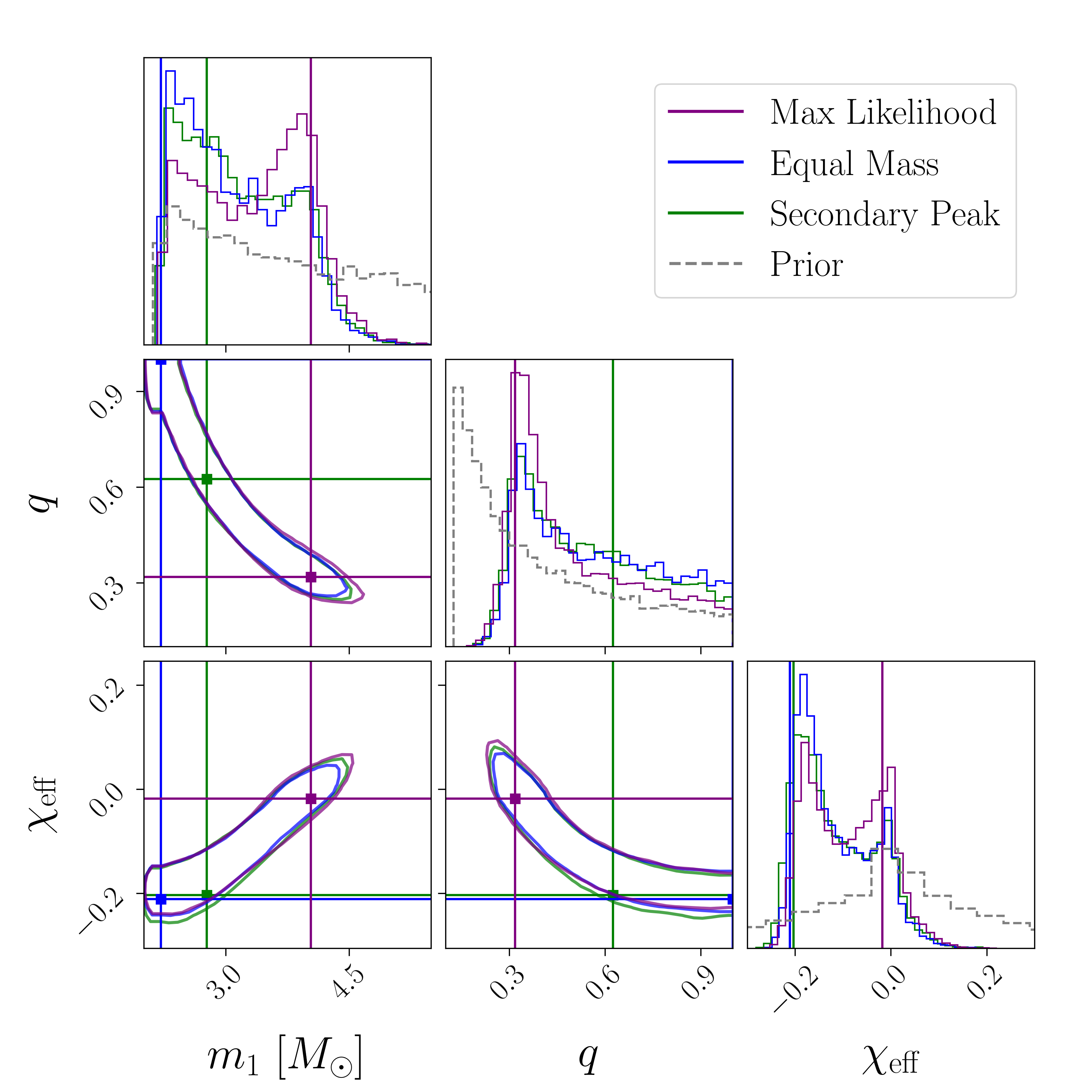} 
\caption{Posterior probability distributions of the source-frame primary mass, mass ratio, and effective inspiral spin inferred from parameter estimation of GW230529-like events simulated with zero noise, using the \texttt{IMRPhenomPv2\_NRTidalv2} waveform model.  
The posteriors for simulated binaries with intrinsic parameters corresponding to the \textsc{Max Likelihood} sample (unequal masses characteristic of a mass-gap \ac{BH} and a \ac{NS}), the \textsc{Equal Mass} sample (characteristic of two heavy \acp{NS}), and the \textsc{Secondary Peak} sample are shown in purple, blue, and green, respectively. The true binary parameter values used for each simulation are shown with solid lines in the corresponding color. The prior distributions are shown as dashed gray lines in the one-dimensional plots. }
\label{fig:varied_sample_corner}
\end{figure}

In Fig.~\ref{fig:varied_sample_corner}, we show the posterior probability distributions of the primary mass, mass ratio, and effective inspiral spin inferred for the three simulated GW230529-like signals.  
We find that all three primary mass distributions exhibit the same two peaks, at $\sim 2.4 ~\Msun$ and $\sim 4.0 ~\Msun$, and the mass ratio distributions all peak at $q\sim0.3$.

These results imply that, given the \ac{SNR} of the event, the low-mass \ac{NSBH} and \ac{BNS} merger signals are difficult to distinguish from each other. 
The posterior distributions are largely influenced by the priors on the specified parameters, which are overwhelming the likelihood and causing this parallel behavior between simulations.
For instance, even though the true value of the mass ratio for the \textsc{Equal Mass} simulation is $q\simeq1$, this region of the parameter space is disfavored by the prior (shown by the dashed line in Fig.~\ref{fig:varied_sample_corner}). 
Therefore, the posterior peaks at an unequal mass ratio, as it does for the other simulations.
Similarly, the $\chi_{\mathrm{eff}}$ prior disfavors large, negative spins and has maximal support for zero spin, so the posteriors on this parameter exhibit bimodalities at both $\chi_{\mathrm{eff}}\approx 0$ (near the true simulated value for the \textsc{Max Likelihood} simulation) and $\chi_{\mathrm{eff}}\approx -0.2$ (near the true value for the \textsc{Equal Mass} and \textsc{Secondary Peak} simulations).

Given the sensitivity of our posterior constraints on the shape of the prior, it may be worthwhile to consider other prior choices, such as population-informed priors like those explored in the original GW230529 analysis~\citep{LIGOScientific:2024elc}, when evaluating similar signals in future work.
Astrophysically motivated priors will also undoubtedly have a significant impact on the inference results of GW230529-like systems; such priors have been shown to lead to different astrophysical interpretations of \ac{GW} systems compared to the broad, uninformative priors typically used in \ac{LVK} analyses~\citep[e.g.,][]{Vitale:2017cfs,Mandel:2020lhv,Zevin:2020gxf,Mandel:2021ewy,Chattopadhyay:2024hsf}. 
However, given the limited observational evidence for compact objects residing in or in close proximity to the lower mass gap, as well as the uncertainties in the \ac{NS} \ac{EOS}, such prior choices should be applied with caution for GW230529-like systems. 
In this work, we thus focus on the standard uninformative priors used in \ac{LVK} analyses to determine which properties intrinsic to the system drive statistical precision rather than achieving this precision through the use of restrictive priors.

\subsection{Varying the primary mass}

\begin{figure*}[t!]
\plotone{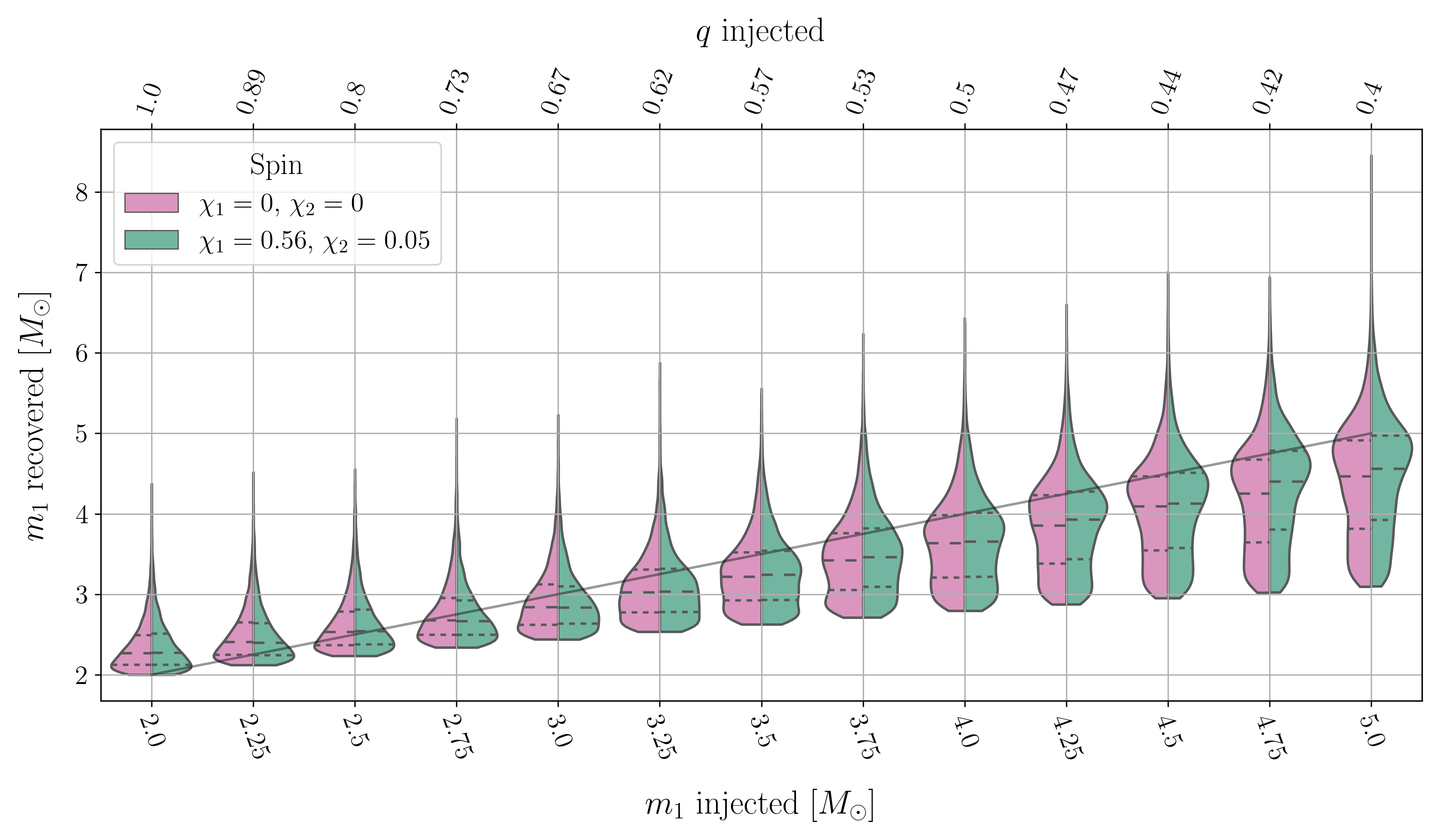} 
\caption{Violin plot showing the recovered posterior probability densities for each of the simulated primary mass values in the detector frame, for both nonspinning (pink) and spinning (teal) systems. The black diagonal line marks the true $m_{1}$ values of the simulations. The true mass ratio values associated with each primary mass are given along the top axis of the plot. }
\label{fig:varied_m1_violin}
\end{figure*}

To further explore the measurability of the primary mass parameter, we broaden our analysis to now include systems beyond only those with exact support in the posterior of GW230529. 
We run simulations over an equally spaced grid of 13 primary mass values between 2 and 5~\Msun (i.e., a simulation at every 0.25~\Msun), while fixing the true value of the secondary mass at 2~\Msun (all values are reported in the detector frame).
By varying the primary mass value, we simultaneously experiment with different values of the mass ratio as well as chirp mass.
Since this parameter is well measured, most of the chirp masses adopted for our simulations are no longer consistent with the inferred chirp mass of GW230529.
Therefore, our goal is not necessarily to simulate a system with all of the properties of GW230529, but instead to explore whether certain values of chirp mass are better measured than others, especially when the component masses are in and around the lower mass gap.

To widen the range of potential chirp mass values, we modify the prior on the detector-frame chirp mass to $\mathcal{M} \in [2.3,  2.8]~\Msun$ for the first seven primary mass values in the grid and $\mathcal{M} \in [1.7, 2.4]~\Msun$ for the remaining six.
The other parameters in the low secondary-spin prior remain the same as in Section~\ref{sec:methods}.
We again use \texttt{IMRPhenomPv2\_NRTidalv2} as our waveform model and a zero-noise realization.
By increasing the luminosity distance as we increase the primary mass, we keep the \ac{SNR} constant at 11.83 to match the \ac{SNR} of the \textsc{Max Likelihood} simulation.
Additionally, we simulate both spinning and nonspinning systems for each value of the primary mass; for the spinning systems, we use the spin parameters of the \textsc{Max Likelihood} sample.

The posterior probability distributions for the recovered primary mass against each of the true detector-frame primary mass values, for systems with and without spin, are shown in Fig.~\ref{fig:varied_m1_violin}.
We find that the uncertainty in the $m_1$ posterior increases monotonically as the primary mass increases (and as the mass ratio decreases); for example, the width of the 90\% credible interval of the posterior probability distribution at $m_1 = 5.0~\Msun$ is $\approx 3$ times the width at $m_1 = 2.0~\Msun$ for nonspinning systems. 
The relative uncertainty for $m_1$, which we define as the width of the 90\% credible interval of the $m_1$ posterior divided by the true value of $m_1$, is $\approx20$\% larger at $m_1 = 5.0~\Msun$ than at $m_1 = 2.0~\Msun$ (again for the nonspinning case). 
However, the relative mass uncertainty for both spinning and nonspinning systems does not increase monotonically over the interval; it decreases first before it increases.
Meanwhile, the relative uncertainty in the chirp mass does increase monotonically over the same interval, and is $\approx2.3$ times greater at the maximum chirp mass value than at the minimum.

The most significant bimodality occurs in the primary mass posteriors for values between 3.5 and 4.0~\Msun, suggesting that parameter estimation for systems in this mass range may lead to larger ambiguities in mass measurements and the inferred nature of the source. 
This could explain some of the uncertainty in the original GW230529 analysis, since these values are squarely in the lower mass gap and comparable to the inferred primary mass value of GW230529.
The distributions for each value of primary mass are very similar regardless of spin, so spin is not a significant contributing factor to the width of the posterior or the bimodality.
Therefore, we conclude that the intrinsic mass parameters of the binary have the greatest impact on the recovery of its primary mass.

\section{Varying the noise realization} \label{sec:noise_realizations}

In order to analyze the effect of detector noise on our parameter estimation results and determine if noise contributed to the uncertainty in the measured posterior distribution of GW230529, we performed 10 simulations with different realizations of Gaussian detector noise added to the signal. 
These simulations were run with each of the sets of intrinsic parameters explored in Section~\ref{sec:sim_posterior_samples}, and used the same waveform model, prior, and single-detector configuration as described in Section~\ref{sec:methods}.

\begin{figure}
\plotone{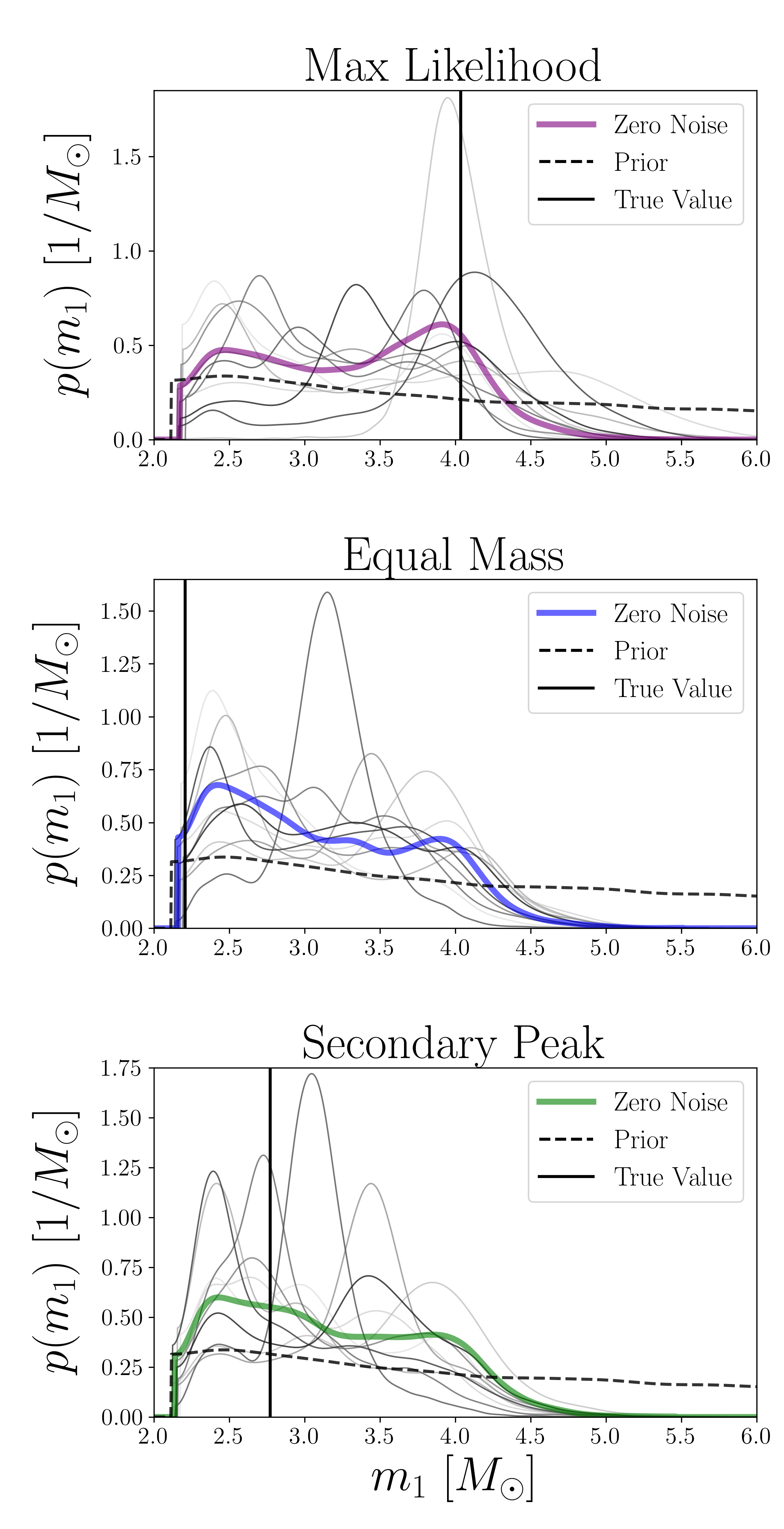} 
\caption{Posterior probability distribution of the source-frame primary mass for each of the three simulated systems corresponding to samples from the GW230529 posterior for 10 different Gaussian noise realizations (gray). The true value for the primary mass is shown by the black vertical line in each plot. We also overlay the zero-noise result from Fig.~\ref{fig:varied_sample_corner} in the colored line, and the prior is shown with a black dashed line. }
\label{fig:varied_noise}
\end{figure}

The primary mass posteriors obtained for these simulations are grouped by the true intrinsic parameters in Fig.~\ref{fig:varied_noise}.
We find that while the particular noise realization does lead to significant differences in the posteriors, the zero-noise results are consistent with the variation in the Gaussian noise results in terms of multimodality and posterior width.
For each set of intrinsic parameters, the majority of realizations of Gaussian noise produced a multimodal $m_{1}$ posterior, so the multimodality observed in zero noise is a common result for our simulated systems despite variations in noise.
The posterior widths of the primary mass vary by only $\sim 10\%$ across different noise realizations in all simulations. 
Therefore, we conclude that the particular noise realization alone was not the reason for the bimodality in the GW230529 primary mass posterior. 

\section{Increasing the Signal-to-noise Ratio} \label{sec:increasing_snr}

\begin{figure*}[t!]
\plotone{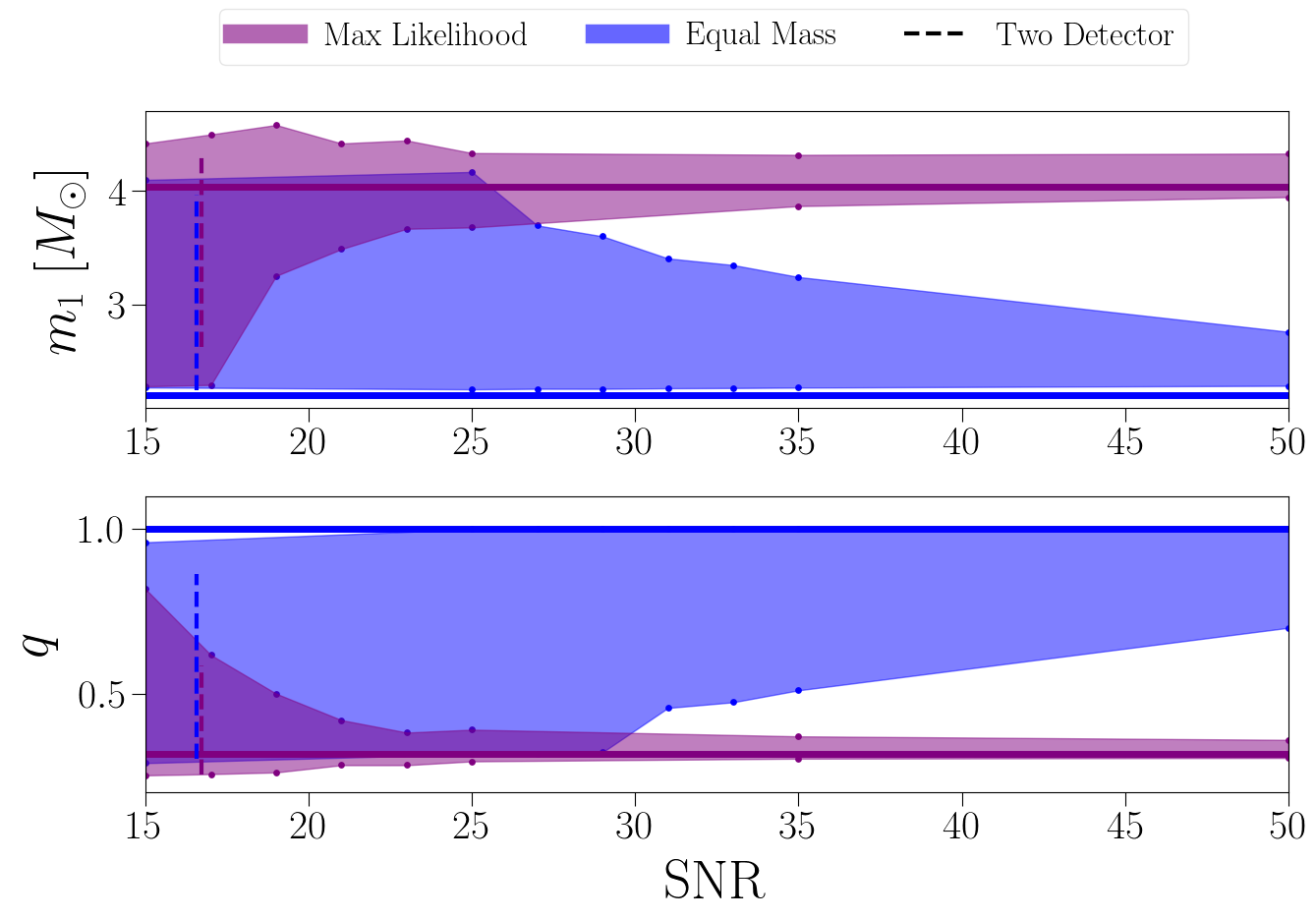} 
\caption{The width of the 90\% credible interval of the posterior probability distributions for primary mass and mass ratio (shaded) vs. simulated \ac{SNR} for the \textsc{Max Likelihood} (purple) and \textsc{Equal Mass} (blue) systems. The horizontal lines represent the true values of each parameter. The vertical dashed lines in the corresponding colors give the width of the 90\% credible intervals when using a two-detector configuration. }
\label{fig:varied_snr}
\end{figure*}

We next investigate how the precision of our measurements would change if the signal were detected with a higher \ac{SNR}. 
We again simulate systems with the \textsc{Max Likelihood} and \textsc{Equal Mass} intrinsic parameters, with one modification: the luminosity distance is adjusted in order to achieve various desired \acp{SNR}. 
Since the $\mathrm{SNR} \approx 11.83$ for our simulation using a zero-noise realization with the \textsc{Max Likelihood}  parameters, we choose new \ac{SNR} values between 15 and 50 to determine how much higher the \ac{SNR} must be to resolve some of the degeneracies in the posterior distributions of the primary mass and mass ratio.
We also perform the analysis using the original luminosity distances listed in Table~\ref{tab:params} and a two-detector configuration (LIGO Livingston and LIGO Hanford). 

Fig.~\ref{fig:varied_snr} shows the width of the $90\%$ credible interval, calculated using the highest posterior density method, of the primary mass and mass ratio posteriors as a function of the simulated \ac{SNR} values. 
As we increase the \ac{SNR} of the signal in our simulations, the width of the posterior decreases until the \textsc{Max Likelihood} and \textsc{Equal Mass} posteriors diverge relative to each other but converge to the true parameter values. 
The behavior depends on the intrinsic parameters used: the $90\%$ credible interval of the \textsc{Max Likelihood} $m_1$ posterior narrows to below 1~\Msun at \ac{SNR}~$\approx~20$, whereas the $90\%$ credible interval of the \textsc{Equal Mass} $m_1$ posterior does not narrow below 1~\Msun until $\mathrm{SNR}\approx~34$.
A higher \ac{SNR} is needed for the likelihood to overcome the prior in the \textsc{Equal Mass} simulation compared to the \textsc{Max Likelihood} simulation, since the priors favor unequal-mass, nonspinning systems over equal-mass, spinning systems.
Consistent with previous results~\citep[e.g.,][]{Arun:2008kb, Vitale:2014mka}, the measurement of spin precession via the effective precessing spin parameter~\citep{Schmidt:2014iyl}, $\chi_p$, also becomes both more accurate and precise with increasing \ac{SNR} and for more unequal mass ratios at the same \ac{SNR}.

Using a two-detector configuration, ${\mathrm{SNR}_{\mathrm{net}}\approx\maxLtwoDetSNR}$ for the \textsc{Max Likelihood} case and ${\mathrm{SNR}_{\mathrm{net}}\approx\eqMasstwoDetSNR}$ for the \textsc{Equal Mass} case. The bounds of the $90\%$ credible interval for $m_1$ using each of these two-detector network \acp{SNR} with corresponding intrinsic parameters are [\twoDetMaxLLowerMass, \twoDetMaxLUpperMass] and [\twoDetEqMassLowerMass, \twoDetEqMassUpperMass], respectively.
Therefore, even the addition of a second detector would not increase the \ac{SNR} enough to distinguish between the two simulated systems, since the 90\% credible intervals of the $m_1$ posteriors using the two-detector network \acp{SNR} largely overlap.


\section{Comparing Waveform Models} \label{sec:waveform_models}

\begin{figure*}
\begin{center}
\includegraphics[scale=0.3,angle=0]{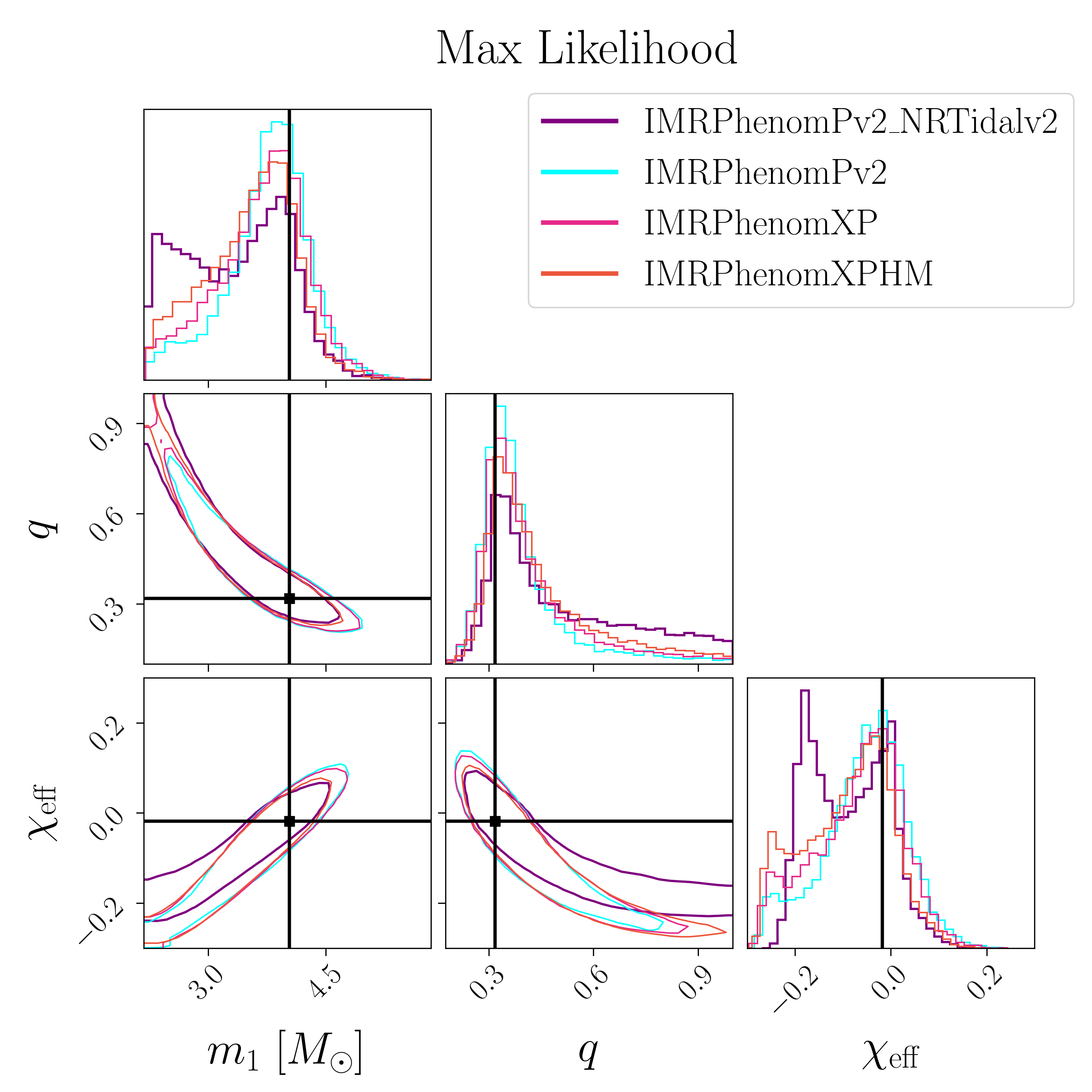} \includegraphics[scale=0.3,angle=0]{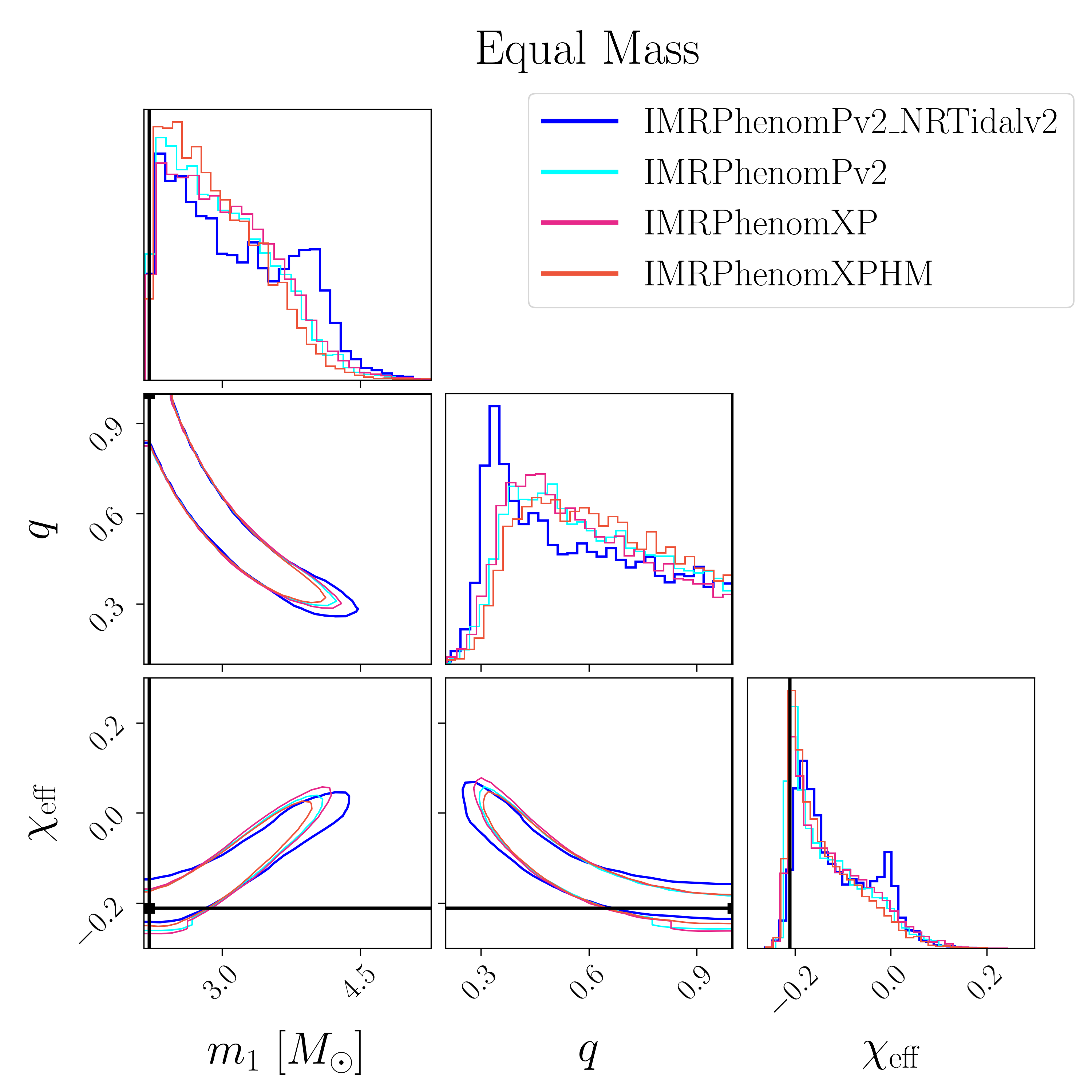} 
\includegraphics[scale=0.3,angle=0]{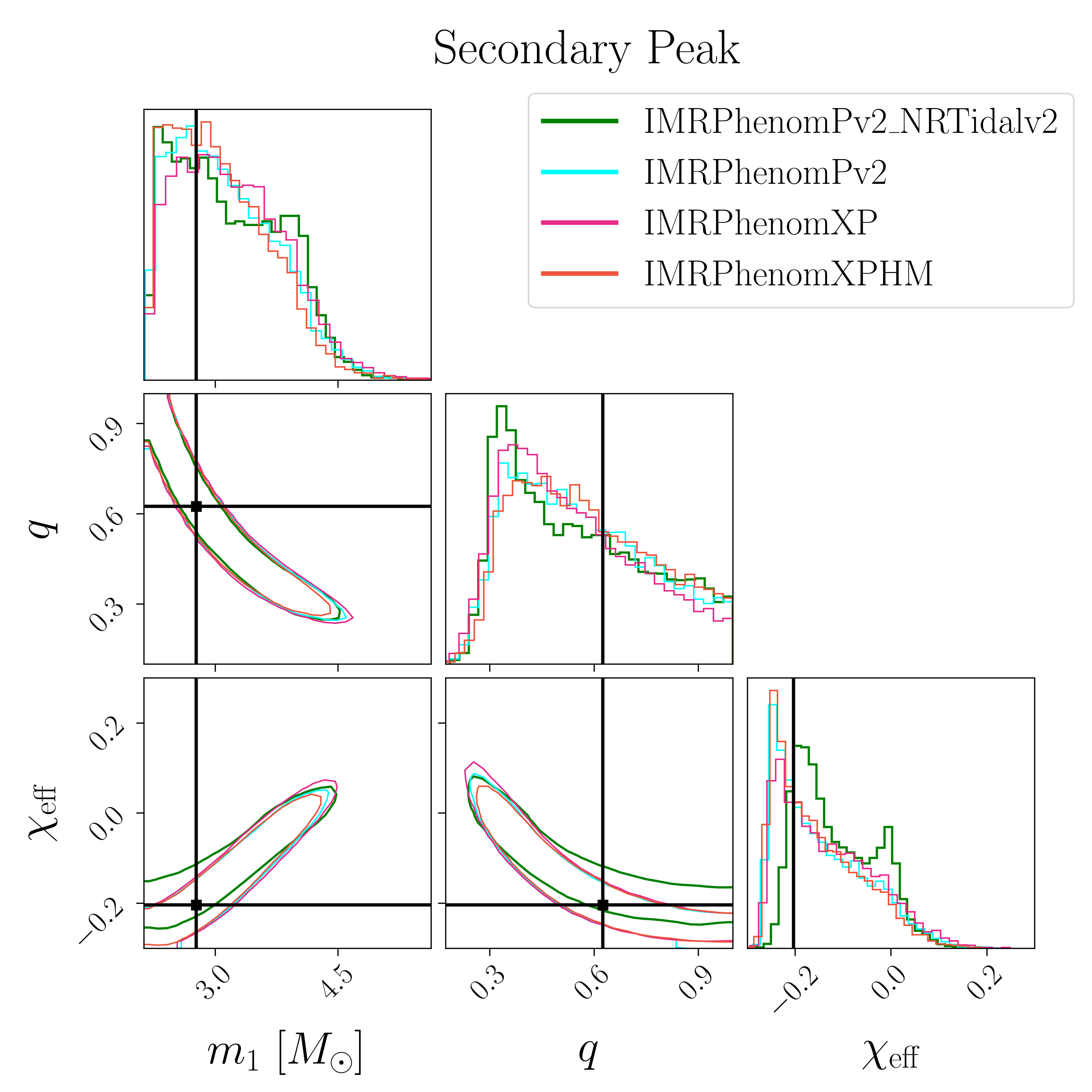}
\caption{Posterior probability distributions of the primary mass, mass ratio, and effective inspiral spin inferred from parameter estimation using different waveform models: \texttt{IMRPhenomPv2\_NRTidalv2} (original colors), \texttt{IMRPhenomPv2} (cyan), \texttt{IMRPhenomXP} (pink), and \texttt{IMRPhenomXPHM} (orange), for each of our three simulations with distinct intrinsic parameters. }
\end{center}
\label{fig:varied_waveform}
\end{figure*}

Finally, for each set of intrinsic parameters, we run simulations using the \texttt{IMRPhenomPv2}, \texttt{IMRPhenomXP}, and \texttt{IMRPhenomXPHM} waveform models to directly compare with our earlier results from Section \ref{sec:intrinsic_params} using the \texttt{IMRPhenomPv2\_NRTidalv2} waveform.
\texttt{IMRPhenomPv2}~\citep{Hannam:2013oca, Khan:2015jqa, Husa:2015iqa} and \texttt{IMRPhenomXP}~\citep{Pratten:2020ceb} model precessing \ac{BBH} systems without higher-order modes; we include them to determine whether the addition of higher-order modes (in the case of \texttt{IMRPhenomXP}) or the elimination of the tidal deformability parameters (in the case of \texttt{IMRPhenomPv2}) is primarily responsible for the differences in the posterior distributions between waveforms. 
In each simulation, we use the same waveform model to generate and recover the signal.
We use a zero-noise realization and the same intrinsic parameters from Table \ref{tab:params}.
For the simulations using the \texttt{IMRPhenomPv2}, \texttt{IMRPhenomXP}, and \texttt{IMRPhenomXPHM} waveform models, we modify the prior range for the secondary-spin magnitude to $\chi_2\le0.99$, since these are \ac{BBH} waveforms not subject to constraints on the \ac{NS} spin. 
We also set the true parameter values of the tidal deformability parameters to zero when using these \ac{BBH} waveforms.

Fig.~\ref{fig:varied_waveform} shows the posteriors obtained through our simulations with each of the waveform models, which can be compared to the real GW230529 posteriors shown in Fig.~\ref{fig:gw230529_pe}.
We note that using the \ac{BBH} waveform models leads to a more precise measurement of the system properties than the \ac{BNS} model \texttt{IMRPhenomPv2\_NRTidalv2} used for our original simulations. 
Especially when using the \textsc{Max Likelihood} parameters, a second peak is most clearly seen in the \texttt{IMRPhenomPv2\_NRTidalv2} posteriors, suggesting that the tidal physics unique to this model and the extra degrees of freedom they contribute to the waveform lead to increased uncertainty.
Because the contribution to the \ac{GW} phase from tidal effects also depends on the masses~\citep{Hinderer:2009ca, Vines:2011ud, DelPozzo:2013ala}, this can introduce additional degeneracies between these parameters.
This is consistent with the results of \cite{Huang:2020pba}, who found that using \texttt{IMRPhenomPv2\_NRTidalv2} for parameter estimation of \ac{NSBH} signals with $q<0.5$ generally leads to the largest statistical uncertainties compared to other waveform models.
These observed statistical uncertainties are a natural consequence of using additional parameters to describe the waveform and not caused by a mismatch between the physics included in the \texttt{IMRPhenomPv2\_NRTidalv2} model and the simulated signal; the inclusion of the tidal deformability parameters reduces the constraints on the measurements of other parameters.

Since the distributions for all of the \ac{BBH} waveform models are similar, we find that the inclusion or omission of radiation from higher-order modes in both simulation and recovery does not significantly impact the posteriors obtained for GW230529-like systems. 
However, for systems like GW230529, higher-order modes are expected to have a greater impact on the waveform than the inclusion of tidal effects, as the former are enhanced in binaries with unequal mass ratios~\citep{Cho:2012ed, OShaughnessy:2014shr, Pratten:2020igi}, whereas the latter primarily affects high frequencies, where current detectors have limited sensitivity. While we use waveforms with tidal effects but without higher-order modes to simulate a subset of the GW230529-like systems explored in this work, the analysis of real \ac{NSBH} signals like GW230529 should prioritize the inclusion of higher-order modes over tidal effects, as waveform models incorporating both simultaneously are still under active development~\cite[e.g.,][]{Abac:2025brd}.
We therefore conclude that the bimodality and ambiguity in the original GW230529 posteriors obtained with the \texttt{IMRPhenomPv2\_NRTidalv2} model (pink lines in Fig.~\ref{fig:gw230529_pe}) primarily result from the extra degrees of freedom from the addition of the tidal deformability parameters in the \ac{BNS} waveform model, although measurements precise enough to distinguish the nature of the source are still possible with this waveform model at higher \ac{SNR} (see Fig.~\ref{fig:varied_snr}).

\section{Conclusions} \label{sec:conclusions}
In this work, we have investigated sources of statistical uncertainty in the measured properties of compact binary coalescences with primary component masses in and around the lower mass gap. 
By performing parameter estimation on simulated \ac{GW} signals with properties similar to GW230529, we examine how the binary parameters, detector noise, \ac{SNR}, and waveform model affect measurements of the properties of the system. 
We focus mainly on the primary mass, mass ratio, and effective inspiral spin parameters to compare the posterior distributions obtained in our systematic simulations. 

The main questions we investigate in this work and our findings are as follows.

\emph{Can we measure the primary mass of a GW230529-like source well enough to distinguish whether it is consistent with a massive \ac{BNS} merger or a \ac{NSBH} system?} 
We cannot confirm whether GW230529 resulted from the merger of two massive \acp{NS} or a standard-mass \ac{NS} and a low-mass \ac{BH}, although the latter option remains most likely, given current knowledge of the \ac{NS} \ac{EOS}. 
We find that we cannot confidently constrain the primary mass parameter for simulated systems like GW230529 to either above or below 3~\Msun, the approximate boundary between \acp{NS} and \acp{BH}, which leads to the ambiguous classification of the primary component.

\emph{How well can we measure the properties of lower-mass-gap systems with different masses?}
The primary mass posterior obtained through parameter estimation is most strongly multimodal for systems with chirp masses similar to the inferred chirp mass of GW230529 ($\sim2.026~\Msun$ in the detector frame).
Consistent with expectations, we find that the variability of the recovered primary mass distribution depends primarily on the true mass parameters and is not strongly affected by whether the system is spinning or nonspinning.
Although we cannot measure the mass precisely enough to determine which type of compact object it is, we can constrain the primary mass for simulated systems similar to GW230529 at the $\approx 36\%$ level.
This supports the conclusions of previous studies, that \acp{BH} and \acp{NS} are hard to distinguish with \ac{GW} observations at low \acp{SNR} when the component masses of the system are inferred to be within or near the lower-mass-gap region of the parameter space~\citep{Hannam:2013uu, Littenberg:2015tpa, Tsokaros:2019lnx, Yang:2017gfb}. 
If we want to confidently identify whether an object in this type of system is a \ac{BH} or a \ac{NS} and reliably measure the properties of the binary, additional information such as electromagnetic observations of the event may be needed, strengthening the motivation for multimessenger follow-up of \ac{GW} sources in this mass range.

\emph{Is detector noise the reason for the ambiguity in the GW230529 parameter estimation results?}
The specific realization of noise in the detector is not the main reason for the ambiguity in the results of the parameter estimation. 
We find that posteriors produced by simulations with no added detector noise, as well as the majority of those obtained using different simulated Gaussian noise realizations, exhibit similar statistical uncertainties and multimodalities as the original posteriors from the GW230529 parameter estimation.

\emph{How loud would a GW230529-like signal have to be to determine the nature of the primary object?}
Increasing the \ac{SNR} quickly improves the precision of our measurements by eliminating the bimodality and reducing the width of the mass posteriors.
Due to the overlapping mass posteriors at low \acp{SNR} for simulated high-mass \ac{BNS} and mass-gap \ac{NSBH} systems, the low \ac{SNR} of the GW230529 signal is the primary reason for the uncertainty in the posterior distributions of the component masses and effective inspiral spin.
In other words, the \ac{SNR} of the GW230529 signal is not high enough to fully overcome the prior on the parameters of interest. 
We find that the properties of unequal-mass systems are better constrained at a given \ac{SNR} than those of equal-mass systems.
While a \ac{SNR}~$\ge~20$ would be needed to constrain the primary mass measurement to within 1~\Msun for unequal-mass systems similar to GW230529, a \ac{SNR}~$\ge~34$ is required to obtain this same constraint in the equal-mass case.


\emph{How does the waveform model used affect how well the true parameters of the simulated signals are recovered?}
The waveform model chosen to simulate signals and perform parameter estimation on the system has a significant impact on the precision of the posteriors. 
Eliminating the effect of tidal deformability of the compact objects by using one of the \ac{BBH} waveform models, such as \texttt{IMRPhenomPv2}, breaks some of the degeneracies in the posterior distribution obtained from analyses with the \ac{BNS} \texttt{IMRPhenomPv2\_NRTidalv2} waveform model. 
Given the high confidence that the GW230529 source contains at least one \ac{NS} based on mass measurements, tidal effects should not be fully eliminated from the analyses of such events. 
However, the inclusion of tidal deformability parameters adds degrees of freedom to the waveform model, which we find account for much of the observed ambiguity at low \ac{SNR} in the \texttt{IMRPhenomPv2\_NRTidalv2} results. We leave the investigation of whether the tidal effects on the primary or secondary compact object dominate the statistical uncertainty in the mass measurements to future work. It may also be worthwhile to simulate and recover the signal with different waveform models to test how this mismodeling of physical effects translates into the observed degeneracies in the posterior. We reserve this more thorough study of waveform systematics for future work as well.

While we find that the ambiguity in the mass measurements is inherent for systems with binary parameters similar to GW230529 detected at comparable \ac{SNR}, our simulations suggest that future observations of these types of systems could deepen our understanding of multiple astrophysical processes. Precise measurements of the parameters of GW230529-like systems and the ability to distinguish \acp{BH} from \acp{NS} in lower-mass-gap sources would have major astrophysical implications for the \ac{NS} \ac{EOS}, supernova mechanisms, compact binary formation channels, and multimessenger prospects for \ac{NSBH} systems.

Posterior samples for our parameter estimation, as well as initialization files for these runs, are available on Zenodo, doi:10.5281/zenodo.18177233~\citep{data_release}.

\vspace{10pt}
This material is based upon work supported by the National Science Foundation under grant No. AST2149425, a Research Experiences for Undergraduates (REU) grant awarded to CIERA at Northwestern University. Any opinions, findings, and conclusions or recommendations expressed in this material are those of the author(s) and do not necessarily reflect the views of the National Science Foundation.
This material is also based upon work supported by NSF's LIGO Laboratory, which is a major facility fully funded by the National Science Foundation.
LIGO was constructed by the California Institute of Technology and Massachusetts Institute of Technology with funding from the National Science Foundation and operates under cooperative agreement PHY-0757058. 
The authors thank Vicky Kalogera for providing guidance and support, and Geraint Pratten for helpful comments on this manuscript. 
J.C. extends thanks to Aaron Geller and Chase Kimball for their leadership in the REU program through which this work originated. M.Z. gratefully acknowledges funding from the Brinson Foundation in support of astrophysics research at the Adler Planetarium. S.B. is supported by NASA through the NASA Hubble Fellowship grant No. HST-HF2-51524.001-A awarded by the Space Telescope Science Institute, which is operated by the Association of Universities for Research in Astronomy, Inc., for NASA, under contract NAS5-26555.
This research was supported in part through the computational resources and staff contributions provided for the Quest high-performance computing facility at Northwestern University, which is jointly supported by the Office of the Provost, the Office for Research, and Northwestern University Information Technology.
The authors are grateful for computational resources provided by the LIGO Laboratory and supported by NSF grant Nos. PHY-0757058 and PHY-0823459. This paper carries LIGO document number LIGO-P2500385.


\bibliography{citations}{}
\bibliographystyle{aasjournal}

\end{document}

%% file: acronyms.tex
\acrodef{GW}{gravitational-wave}
\acrodef{LVK}{LIGO-Virgo-KAGRA}
\acrodef{SNR}{signal-to-noise ratio}
\acrodef{EOS}{equation of state}
\acrodef{BBH}{binary black hole}
\acrodef{BNS}{binary neutron star}
\acrodef{NSBH}{neutron star--black hole}
\acrodef{BH}{black hole}
\acrodef{NS}{neutron star}
\acrodef{O4}{fourth observing run}

\acrodef{L1}{LIGO Livingston}
\acrodef{H1}{LIGO Hanford}

%% file: macros.tex
\newcommand{\originalEventSNR}{11.4}
\newcommand{\maxLtwoDetSNR}{16.72}
\newcommand{\eqMasstwoDetSNR}{16.57}
\newcommand{\twoDetMaxLLowerMass}{2.63}
\newcommand{\twoDetMaxLUpperMass}{4.34}
\newcommand{\twoDetEqMassLowerMass}{2.25}
\newcommand{\twoDetEqMassUpperMass}{3.97}

\newcommand{\minChirpPrior}{2.0214}
\newcommand{\maxChirpPrior}{2.0331}

%% file: parameters_table.tex
\begin{deluxetable*}{c c c c c c c c c}[t]
\label{tab:params}
\tablecaption{Parameter values of the GW230529 posterior samples chosen for our analyses.}
\setlength{\tabcolsep}{10pt}
\renewcommand{\arraystretch}{1.4}
\tablehead{
\colhead{Sample Name} & \colhead{$m_1/M_\odot$} & \colhead{$m_2/M_\odot$} & \colhead{$q$} & \colhead{$\chi_{\mathrm{eff}}$} & \colhead{$D_\mathrm{L}$/Mpc} & \colhead{$\Lambda_1$} & \colhead{$\Lambda_2$} & \colhead{SNR}
}
\startdata
\textsc{Max Likelihood} & 4.03 & 1.28 & 0.32 & -0.02 & 259.80 & 173.25 & 94.28 & 11.83 \\
\textsc{Equal Mass} & 2.21 & 2.21 & 1.00 & -0.21 & 250.72 & 621.12 & 1627.91 & 11.38 \\
\textsc{Secondary Peak} & 2.77 & 1.73 & 0.63 & -0.20 & 317.39 & 41.69 & 1356.19 & 10.47 \\
\enddata
\vspace{4pt}
\tablecomments{$m_1$ and $m_2$ represent the source-frame masses. The optimal SNR listed is for a single-detector configuration (LIGO Livingston only).}
\end{deluxetable*}